\documentclass{article}

\usepackage[preprint,nonatbib]{neurips_data_2024}

\usepackage[utf8]{inputenc} 
\usepackage[T1]{fontenc}    
\usepackage{hyperref}       
\usepackage{url}            
\usepackage{booktabs}       
\usepackage{amsfonts}       
\usepackage{amsmath}
\usepackage{nicefrac}       
\usepackage{microtype}      
\usepackage{xcolor}         
\usepackage{graphicx}
\usepackage{lipsum}
\usepackage{listings}
\usepackage{cprotect}

\title{Shadow and Light: Digitally Reconstructed Radiographs for Disease Classification}

\author{%
Benjamin Hou$^{1*}$ \quad 
Qingqing Zhu$^{2}$ \quad 
Tejas Sudarshan Mathai$^1$ \quad 
Qiao Jin$^2$  \\
\textbf{Zhiyong Lu}$^2$ \quad 
\textbf{Ronald M. Summers}$^1$ \\
$^1$Imaging Biomarkers and Computer-Aided Diagnosis Laboratory, Clinical Center \\ 
$^2$National Center for Biotechnology Information, National Library of Medicine \\
\texttt{\{benjamin.hou, qingqing.zhu, tejas.mathai, qiao.jin\}@nih.gov}\\
\texttt{luzh@ncbi.nlm.nih.gov, rsummers@cc.nih.gov}
}

\begin{document}

\maketitle

\begin{abstract}
In this paper, we introduce DRR-RATE, a large-scale synthetic chest X-ray dataset derived from the recently released CT-RATE dataset. DRR-RATE comprises of 50,188 frontal Digitally Reconstructed Radiographs (DRRs) from 21,304 unique patients. Each image is paired with a corresponding radiology text report and binary labels for 18 pathology classes. Given the controllable nature of DRR generation, it facilitates the inclusion of lateral view images and images from any desired viewing position. This opens up avenues for research into new and novel multimodal applications involving paired CT, X-ray images from various views, text, and binary labels. We demonstrate the applicability of DRR-RATE alongside existing large-scale chest X-ray resources, notably the CheXpert dataset and CheXnet model. Experiments demonstrate that CheXnet, when trained and tested on the DRR-RATE dataset, achieves sufficient to high AUC scores for the six common pathologies cited in common literature: Atelectasis, Cardiomegaly, Consolidation, Lung Lesion, Lung Opacity, and Pleural Effusion. Additionally, CheXnet trained on the CheXpert dataset can accurately identify several pathologies, even when operating out of distribution. This confirms that the generated DRR images effectively capture the essential pathology features from CT images. The dataset and labels are publicly accessible at \url{https://huggingface.co/datasets/farrell236/DRR-RATE}.

\end{abstract}

\section{Introduction}

Chest X-ray imaging is a cornerstone diagnostic tool in clinical practice, providing crucial insights into a wide range of pulmonary and cardiac conditions. Chest X-rays facilitate the diagnosis of various abnormalities by radiologists, such as pneumonia, pneumothoraces and lung lesions. It is a common imaging modality used in resource-constrained settings and often requires compression for use in various applications, such as telemedicine. Therefore, it is often important to be able to potentially diagnose different conditions in Chest X-rays of varying image quality. With the increasing popularity and accessibility of artificial intelligence (AI), significant advancements have been made in the automated analysis of medical images, especially in automated diagnosis, classification, and segmentation. These developments in computational diagnostics heavily rely on the availability of large, well-curated datasets capable of training, validating, and testing these advanced algorithms.

In recent literature, there has been a surge in the development of multimodal models, with prominent examples including Large Language Models (LLMs) and Vision-Based Large Language Models. These models depend heavily on large and diverse datasets to enhance their performance and accuracy. In this work, we employ a technique known as Digitally Reconstructed Radiography (DRR) to synthetically generate X-ray images from a recently released CT dataset \cite{hamamci2024foundation}. The CT-RATE dataset is not only rich with binary labels, but also includes detailed radiological text reports, providing a valuable resource for training classifiers to diagnose diseases. By utilizing DRR, we create a paired dataset that bridges X-ray and CT imagery, potentially opening new avenues for advanced multimodal applications that leverage both visual and textual data to improve diagnostic and predictive capabilities in the medical field.

\subsection{Background}

\begin{figure}[h]
    \centering
    \includegraphics[trim={2cm 0 1cm 0},clip,height=3.2cm]{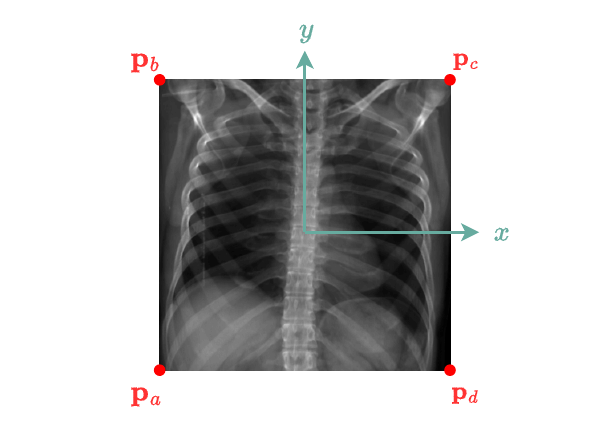}
    \includegraphics[height=3.2cm]{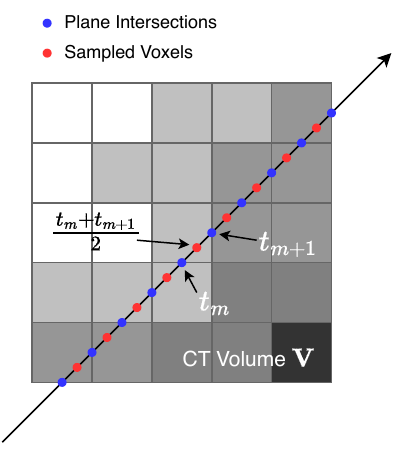}
    \includegraphics[height=3.2cm]{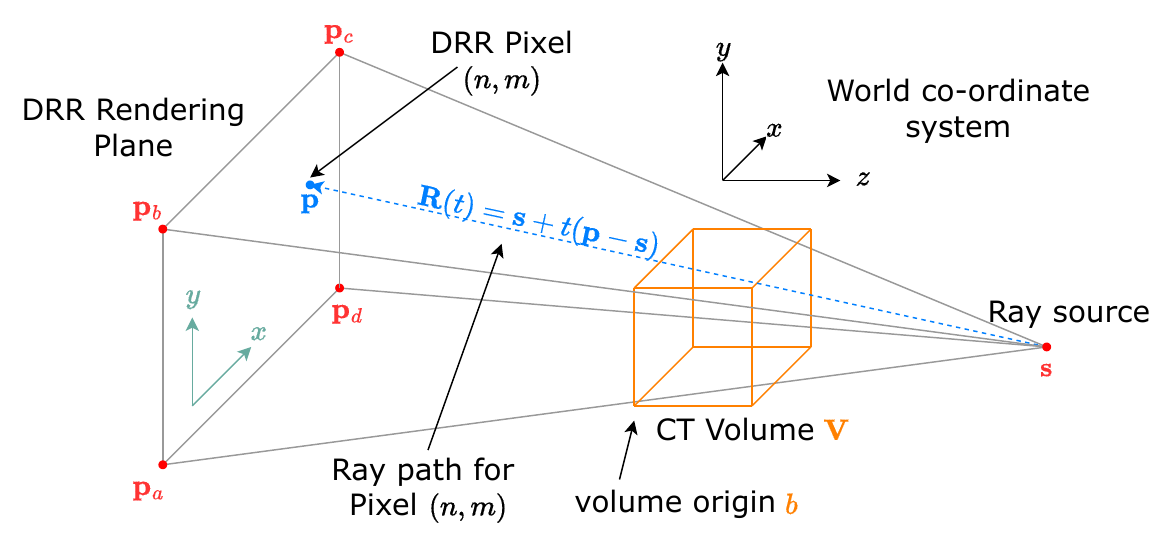}
    \caption{DRR Synthesis Overview: (right) An idealized model of a projectional radiography imaging system. X-ray beams, $\mathbf{R}(t)$, originating from a fixed position, $\mathbf{s}$, and with constant energy diminish in intensity as they travel through the CT volume, $\mathbf{V}$. The attenuation, with respect to volume, is measured when the X-rays reach a point, $\mathbf{p}$, on the detector plane, resulting in the production of the DRR. (middle) Siddon's method of voxel interpolation. The pixel value at $\mathbf{p}$ represents the weighted average of the intensities of all voxels intersected by the ray. The weight corresponds to the length of the line segment passing through each voxel. $\frac{t_{m+1} - t_{m}}{2}$ denotes the midpoint, while $t_{m}$ and $t_{m+1}$ denote the boundary intersections of the voxel. (left) A rendered DRR from a chest CT volume.
}
    \label{fig:drr-overview}
\end{figure}

Digitally Reconstructed Radiographs (DRRs) are synthetic X-ray images created from computed tomography (CT) data. Unlike conventional radiographs, DRRs offer more controlled and reproducible imaging conditions. This is achieved through ray tracing, a technique that simulates the path of X-rays as they pass through the CT volume, as shown in Fig. \ref{fig:drr-overview}. Ray tracing algorithms compute the interaction of rays with the volumetric data by tracing the paths of individual rays from the X-ray source, through the patient model, and onto the detector plane. Each ray accumulates intensity values based on the attenuation coefficients of the tissues it traverses, which are determined by the densities and atomic compositions of those tissues in the volume. This accumulation reflects the integral of the linear attenuation coefficient along the ray path, analogous to the Beer-Lambert law in optics. The result is a grayscale image where each pixel intensity corresponds to the projected X-ray absorption, providing a valuable tool for applications such as radiation therapy planning, where accurate dose calculations and patient positioning are crucial.

Digitally Reconstructed Radiographs (DRRs) are prominently featured for applications in radiation therapy planning, surgical preparation, educational purposes, algorithm development, and quality assurance. In radiation therapy, DRRs are essential for simulating and verifying targeted radiation doses \cite{galvin1995use,Stanley_1999,bastida2011drr}. By aligning the DRR with the actual patient X-ray taken during treatment, clinicians can ensure that the radiation beams are accurately targeted to the tumor, minimizing exposure to surrounding healthy tissues. DRRs also play a pivotal role in the 2D-3D registration process \cite{Wu2010,munbodh2006automated}, where they help align 2D X-ray images with 3D CT scans, e.g., in C-arm X-ray imaging \cite{hou2017predicting}. Created from volumetric CT scans, DRRs closely mimic the 2D X-ray images. This alignment is crucial in specialized procedures like orthopedic surgery or vascular treatments, where accurate matching of 3D CT models with 2D imaging is essential for effective surgical planning. Educational resources leverage DRRs to enhance learning among medical professionals by providing realistic representations of various conditions \cite{sturgeon2015improved,cosson2020geometric}. Furthermore, researchers are also advancing the DRR generation method, e.g., accelerating the generation process and/or enhancing image quality \cite{dorgham2012gpu,bhat2017accelerated,dhont2020realdrr,gopalakrishnan2022fast}.

\subsection{Related Works}

Several notable large-scale public datasets of chest X-rays (CXR) have been released in recent literature, which are pivotal in advancing research and development in medical imaging. Table \ref{tab:common-cxr} provides a detailed overview of key chest X-ray datasets utilized in research over the past decade.

\begin{table}[!ht]
\centering
\caption{Overview of key chest X-ray datasets in recent literature.}
* Labeled by NLP algorithm | † Labeled by radiologist
\begin{tabular}{@{}lccccc@{}}
\toprule
~ Dataset       & Release Year & \# Findings & \# Samples & \# Patients & Has Text ~\\ 
\midrule
~ JRST \cite{shiraishi2000development}        
                & 2000         & 1           & 247*       & 247         & No          ~\\
~ MC \cite{jaeger2014two}          
                & 2014         & 1           & 138*       & 138         & Yes         ~\\
~ Shenzhen \cite{jaeger2014two}    
                & 2014         & 1           & 662*       & 662         & Yes         ~\\
~ OpenI \cite{demner2016preparing}        
                & 2016         & 10          & 8,121*     & 3,955       & Yes         ~\\
~ ChestX-ray8 \cite{wang2017chestx}
                & 2017         & 8           & 108,948†   & 32,717      & No          ~\\
~ ChestX-ray14 \cite{wang2017chestx}
                & 2017         & 14          & 112,120†   & 30,805      & No          ~\\
~ CheXpert \cite{irvin2019chexpert}     
                & 2019         & 14          & 224,316†   & 65,240      & No          ~\\
~ PadChest \cite{bustos2020padchest}     
                & 2019         & 193         & 160,868*†  & 67,000      & Yes         ~\\
~ MIMIC-CXR \cite{johnson2019mimic}    
                & 2019         & 14          & 377,110†   & 65,079      & Yes         ~\\
~ VinDr-CXR \cite{nguyen2022vindr}    
                & 2020         & 28          & 18,000*    & ---         & No          ~\\ 
\bottomrule
\end{tabular}
\label{tab:common-cxr}
\end{table}

Among them are ChestX-ray8 and its extended version, ChestX-ray14 \cite{wang2017chestx}, released by the US National Institutes of Health (NIH), that encompasses over 112,000 CXR scans from 30,805 unique individuals. The data was sourced from the NIH Clinical Center, and disease labels were extracted using natural language processing (NLP) techniques from associated radiological reports.

CheXpert \cite{irvin2019chexpert}, another significant dataset, includes 224,316 chest radiographs from 65,240 patients who underwent radiographic examinations at Stanford Health Care from October 2002 to July 2017. The data was labeled using NLP techniques (primarily by the CheXpert labeler), and supplemented with evaluation sets that were manually labeled by radiologists as reference standards.

PadChest \cite{bustos2020padchest} is another large-scale, high-resolution labeled chest X-ray dataset designed for the automated exploration of medical images along with their associated reports. The dataset includes more than 160,000 chest X-ray images obtained from 67,000 patients. These images were interpreted and reported by radiologists at Hospital San Juan Hospital in Spain from 2009 to 2017. Reports associated with the X-ray images were labeled with 174 different radiographic findings, 19 differential diagnoses, and 104 anatomic locations. These labels are organized in a hierarchical taxonomy and mapped onto the standard Unified Medical Language System (UMLS) terminology. 

Other notable datasets include the MIMIC-CXR \cite{johnson2019mimic} and VinDr-CXR \cite{nguyen2022vindr}. The MIMIC-CXR is a large, publicly available resource of chest radiographs with structured labels. It contains 371,920 images corresponding to 224,548 radiographic studies conducted at the Beth Israel Deaconess Medical Center in Boston, MA. The VinDr-CXR dataset is an extensive, open collection of chest X-rays, complete with radiologist annotations. It comprises over 100,000 raw images in DICOM format, retrospectively collected from two major hospitals in Vietnam—Hospital 108 and Hanoi Medical University Hospital. This dataset includes 18,000 postero-anterior (PA) view chest X-ray scans, annotated by a team of 17 radiologists, each with at least eight years of experience. Taken all together, these datasets facilitate a wide range of research into disease detection and development of artificial intelligence applications in radiology and beyond.

\section{DRR-RATE Dataset}
\label{sec:drr-rate}

The DRR-RATE dataset proposed in this work is built upon the recently released CT-RATE \cite{hamamci2024foundation} dataset, which comprises 25,692 non-contrast chest CT volumes from 21,304 unique patients. Each study is accompanied by a corresponding radiology text report and binary labels for 18 pathology classes. The dataset has been expanded to 50,188 volumes through the modification of the reconstruction matrix extracted from the raw DICOM study. As the dataset was already anonymized, compliance with the Health Insurance Portability and Accountability Act (HIPAA) was ensured, and the requirement for informed consent was waived. CT-RATE is published under the Creative Commons Attribution-NonCommercial-ShareAlike (CC BY-NC-SA) license. In accordance with this license, we provide appropriate credit to the original creator, do not use the material for commercial purposes, and distribute any derivatives under the same license. The DRR-RATE dataset is publicly accessible at \url{https://huggingface.co/datasets/farrell236/DRR-RATE}.

\subsection{Patient Statistics}
\begin{figure}[!ht]
\centering
\includegraphics[height=3.6cm]{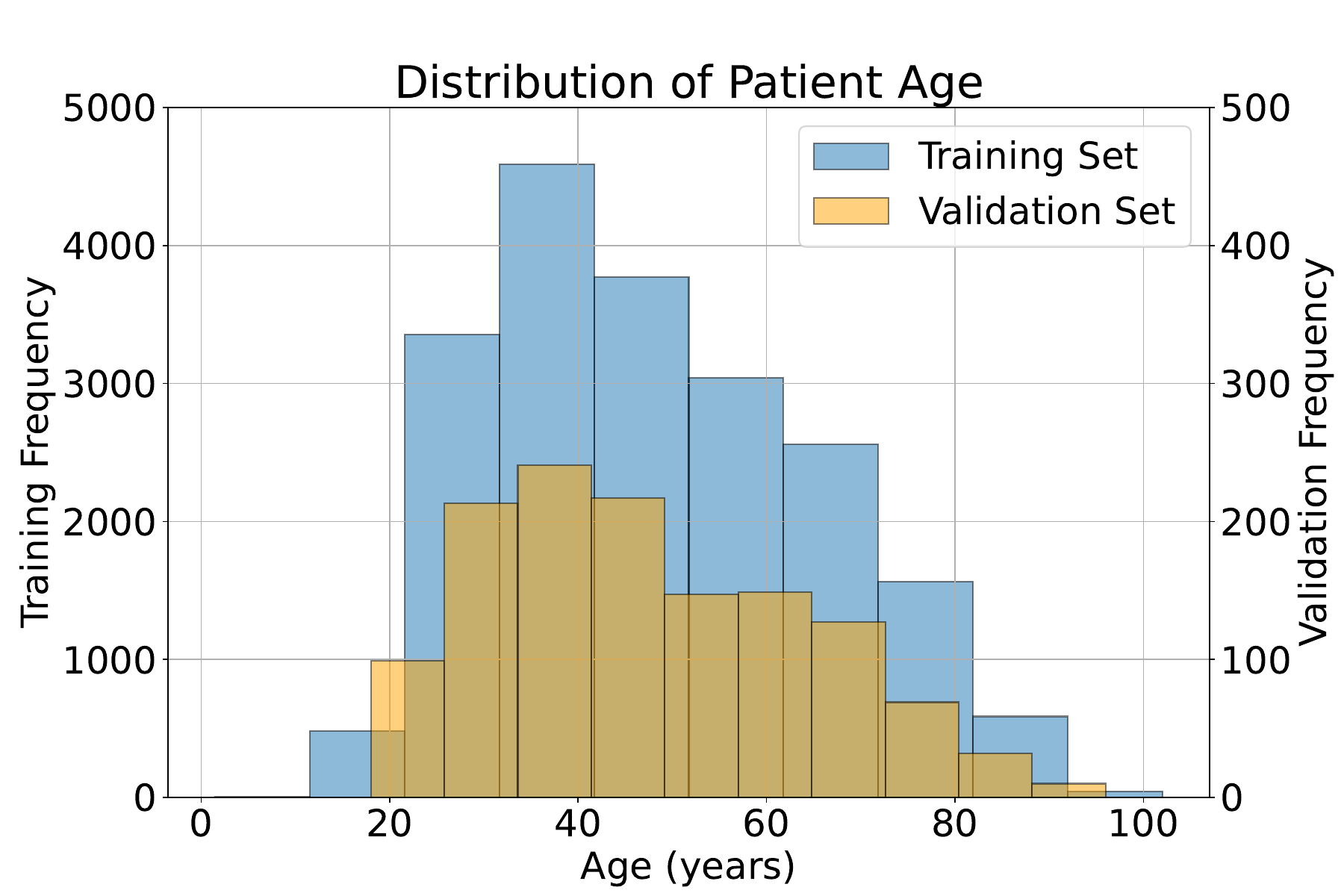} \hspace{2mm}
\includegraphics[height=3.6cm]{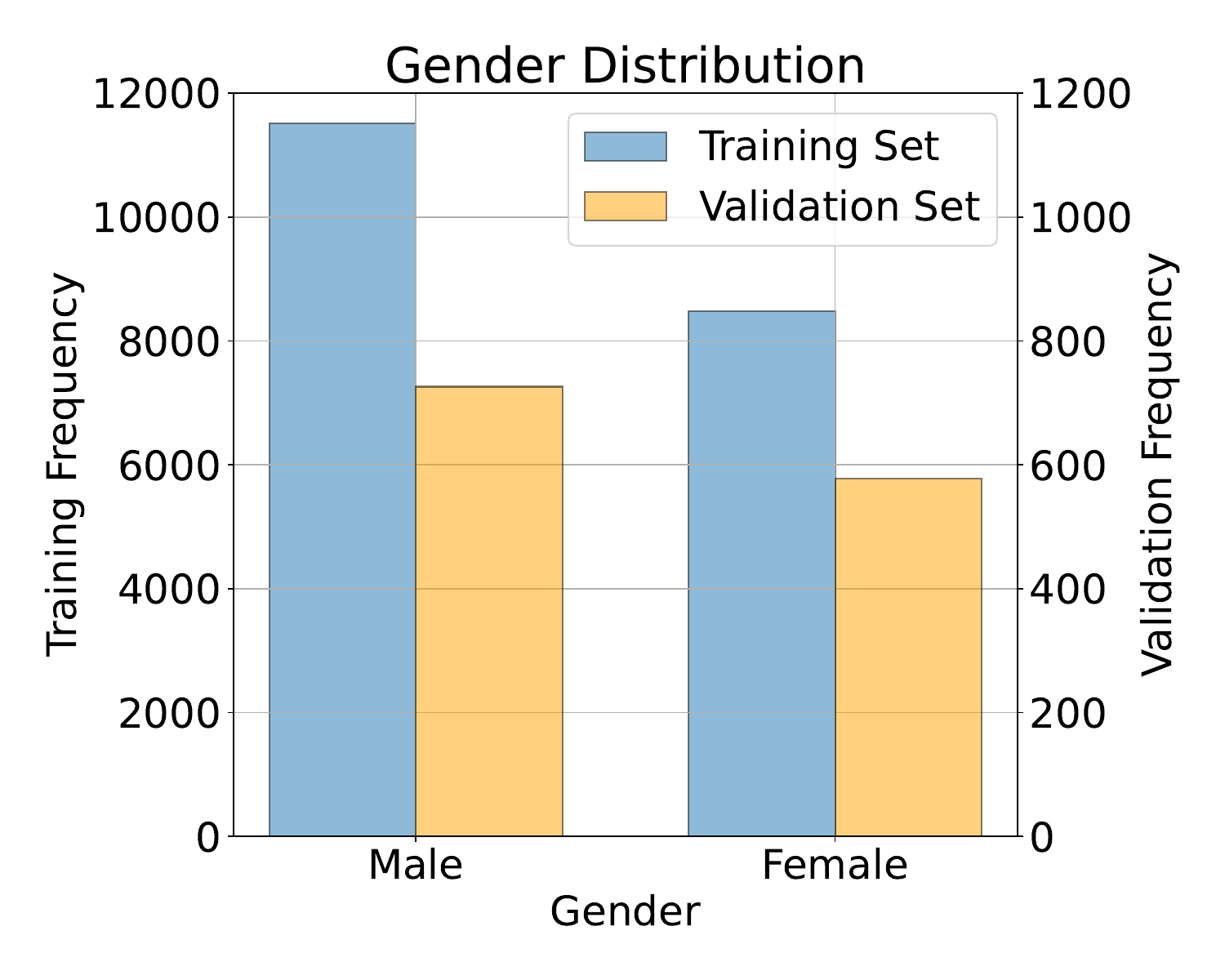}
\caption{Histogram of patient age distribution and gender count}
\label{fig:patient-stats}
\end{figure}

The DRR-RATE dataset is split into training and validation subsets identically to the CT-RATE dataset, and it is utilized to assess classifiers trained on DRR images. There is no official dedicated test set in the CT-RATE dataset. The training set consists of \( N = 19,995 \) recorded patients between ages 1.5 - 102 years. The mean age is \( \mu = 47.85 \) years with a standard deviation of \( \sigma = 17.10 \). The median age is 45, with the first quartile (\( Q1 \)) at 34 and the third quartile (\( Q3 \)) at 61. There are 11516 recorded male patients and 8481 recorded female patients. The validation set consists of \( N = 1,304 \) recorded patients with ages ranging from 18 to 96 years. The mean age is \( \mu = 47.38 \) years with a standard deviation of \( \sigma = 16.73 \). The median age is 45, with the first quartile (\( Q1 \)) at 34 and the third quartile (\( Q3 \)) at 60. There are 726 recorded male patients and 578 recorded female patients. Fig. \ref{fig:patient-stats} details the distribution of patient ages and gender count.

\subsection{DRR Creation}

DRR images are produced by projecting CT volumetric data onto a two-dimensional plane, mimicking conventional X-ray radiographs. Let $\mathbf{s} \in \mathbb{R}^3$ be the source point of the ray in 3D space, and $\mathbf{p} \in \mathbb{R}^3$ be the endpoint of the ray on the detector. The parametric equation of the ray can be expressed as $\mathbf{R}(t) = \mathbf{s} + t (\mathbf{p} - \mathbf{s})$, where it originates from $\mathbf{s}$ (at $t=0$), passes through the imaged volume, and hits the detector plane at $\mathbf{p}$ (at $t=1$). The total energy attenuation of the X-ray beam by the time it reaches the detector is given by the equation:

\begin{equation}
    \mathbf{E}(\mathbf{R}) = {||\mathbf{p} - \mathbf{s}||}_{2} 
    \int_{0}^{1} \mathbf{V}(\mathbf{s} + t (\mathbf{p} - \mathbf{s})) dt
\end{equation}

where $\mathbf{V} \in \mathbb{R}^{3\times3}$ represents the attenuation coefficient at points within the CT volume, and ${||\mathbf{p} - \mathbf{s}||}_{2}$ scales the integral to account for the actual physical distance from the source to the detector, effectively considering the path length of the X-ray. Since the CT volume is composed of discrete voxels, the continuous integral in the equation must be adapted to this discrete structure:

\begin{equation}
    \mathbf{E}(\mathbf{R}) = {||\mathbf{p} - \mathbf{s}||}_{2}
    \sum_{m=0}^{M-1} (t_{m+1} - t_{m})
    \mathbf{V}[\mathbf{s} + \frac{t_{m+1} - t_{m}}{2} (\mathbf{p} - \mathbf{s})]
\end{equation}

where \( t_{m} \) and \( t_{m+1} \) parameterize the points where the ray \( \mathbf{R} \) intersects consecutive voxel boundaries within the volume \( \mathbf{V} \). The term \( (t_{m+1} - t_m) \) represents the length of the ray's segment that crosses through a voxel. The expression \( \frac{t_m + t_{m+1}}{2} \) calculates the midpoint of this segment, identifying the central point within the voxel traversed by the ray. This midpoint is then used to determine the attenuation coefficient at that specific location along the ray.
 
A well-established method for generating DRRs is the Siddon-Jacobs ray tracing algorithm \cite{siddonjacobs}, which is available in the Insight Toolkit (ITK) package v4.x or higher \cite{Wu2010, 10.3389/fninf.2014.00013, midas-journal-784}. The ray intersects the voxel grid at discrete points, and the algorithm optimizes the intersection of rays with the voxels, significantly reducing computational load and improving image accuracy. The essential part of the algorithm involves finding $\{t_m\}_{m=0}^{M}$ values where the ray crosses the planes defining the voxels in the volume $\mathbf{V}$. For a voxel grid defined by planes perpendicular to the $x$, $y$, and $z$-axes at regular intervals, the intersection points $t_x$, $t_y$, and $t_z$ are calculated as follows:

\begin{equation}
    t_x(i) = \frac{b_{x} + i\Delta\mathbf{X} - \mathbf{s}_{x}}
    {\mathbf{p}_{x} - \mathbf{s}_{x}} \quad\quad
    i \in \{0, 1, 2, \ldots, N_x\}
\end{equation}

with similar expressions for $t_y(j)$ and $t_z(k)$. \((b_x, b_y, b_z)\) are the base coordinates of the voxel grid, marking the origin point of the volume \( \mathbf{V} \). The integers \( (i, j, k) \) serve as indices for the planes along these axes, with \( i \), \( j \), and \( k \) taking values from 0 up to \( N_x \), \( N_y \), and \( N_z \), which are the dimensions of the CT volume. The terms \( (\Delta \mathbf{X}, \Delta \mathbf{Y}, \Delta \mathbf{Z}) \) denote the intervals between planes on the respective axes, setting the size and spacing of the voxels and hence determining the grid's resolution and scale. 

The values of \( t \) are calculated for all intersections between the ray \( R \) and the planes along the \( x \), \( y \), and \( z \)-axes, retaining only those intersections that occur within the volume \( V \). The algorithm iterates through the sorted list of \( t \)-values to calculate \( \mathbf{E}(\mathbf{R}) \), which corresponds to the intensity of each pixel in the synthesized DRR. This process typically involves interpolation or accumulation of voxel contributions, based on the material properties or density. Note that the creation of DRRs does not include reflection, refraction, and scattering patterns found in real X-ray systems; however, this approximation has proven sufficient in many published applications as discussed above.


\subsection{DRR Siddon-Jacobs Ray Tracing}
\label{sec:siddonjacobs}

To generate DRRs of the CT-RATE dataset, the binary tool ``\texttt{./getDRRSiddonJacobsRayTracing}'' \cite{Wu2010,midas-journal-784} was utilized. Most parameters were kept at their default settings, with the exception of adjusting the volume by 300 mm along the y-axis to enhance the field of view. A threshold cutoff was also set at -100 Hounsfield units (HU), which primarily aids in differentiating air from more radiologically significant tissues. This cutoff enhances image clarity by focusing on tissues that meaningfully affect x-ray absorption, thereby reducing noise and highlighting critical structures. To create lateral view images, the volume was rotated 90 degrees counterclockwise around the z-axis. This yields 47149 executions for the training set and 3039 executions for the validation set, the following command was run for all volumes:

\begin{verbatim}
    user@machine:~$ ./getDRRSiddonJacobsRayTracing input_volume.nii.gz \
        -o output_drr.png \
        -threshold -100 \
        -t 0 300 0 \
        -rz -90  # if LATERAL view
\end{verbatim}

Figure \ref{fig:example} below presents a synthesized DRR generated from a NIfTI volume, where features such as cardiomegaly are prominently displayed. Additionally, other DRR images clearly show signs of consolidation and central bronchiectasis. However, it is important to note that the accompanying text reports were atypical and translated by the authors of the CT-RATE dataset from another language (Turkish) into English, which may affect their readability and clarity.  


\begin{figure}[!ht]
\centering
\begin{minipage}{.25\textwidth}
\centering
\includegraphics[height=3cm]{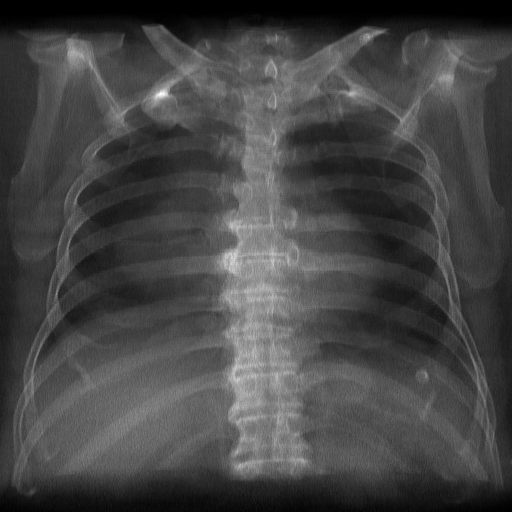} \\
\small 
DRR Image
\label{fig:prob1_6_2}
\end{minipage}%
\begin{minipage}{0.7\textwidth}
\scriptsize
\textbf{Labels:} \verb|{0,0,1,1,0,0,0,0,1,0,0,0,0,0,0,0,0,0}| \\
\textbf{Findings:} Trachea, both main bronchi are open. Heart size increased. Pericardial effusion reaching 2 cm in its thickest part is observed in the pericardial area. Evaluation of mediastinal vascular structures is suboptimal because the examination is unenhanced. As far as can be observed, mediastinal vascular structures were evaluated as normal. Thoracic esophagus calibration was normal and no significant tumoral wall thickening was detected. No lymphadenopathy was detected in the mediastinal area in pathological size and appearance. When examined in the lung parenchyma window; Linear atelectasis areas are observed in the posterobasal sections of both lungs. The upper abdominal organs included in the examination have a natural appearance. Degenerative changes are observed in the bones. No fracture, lytic or destructive lesion was observed. There are extensive osteophytic taperings at the anterior vertebral corners and tend to coalesce.
\end{minipage}
\normalsize
\cprotect\caption{Example DRR image created from \verb|valid_94_a_2.nii.gz|. The binary labels are in the following order: Medical material, Arterial wall calcification, Cardiomegaly, Pericardial effusion, Coronary artery wall calcification, Hiatal hernia, Lymphadenopathy, Emphysema, Atelectasis, Lung nodule, Lung opacity, Pulmonary fibrotic sequela, Pleural effusion, Mosaic attenuation pattern, Peribronchial thickening, Consolidation, Bronchiectasis, and Interlobular septal thickening.
}
\label{fig:example}
\end{figure}

\subsection{DRR Labels}

The radiology text reports, originally in Turkish, were translated into English using an AI-based method and categorized into four groups: Clinical Information, Technique, Findings, and Impression. A text classifier, part of the CT-CLIP project, utilized these translated texts to perform binary classification. The coding and implementation for this classifier has been made publicly available, and functions as an independent application or model. 

The data used for training the model was also sourced from the hospital's PACS system. Although it is not identical to the data released in the CT-RATE dataset, there is considerable overlap, with most of the data also appearing in CT-RATE. A notable strength of this dataset is that all 1000 text label pairs has been validated by a clinician or radiologist from the team. A mapping between the datasets enables the identification of series in CT-RATE whose labels are also clinically validated. These identified series can now be effectively utilized as a dedicated test set, enhancing the robustness of future validations. Table \ref{tab:label-dist} details the available label classes and their distribution.

\begin{table}[!ht]
\centering
\caption{Distribution and Proportion of Training to Validation Labels for CT/DRR-RATE. Note: The data is derived from published CSV label file and has not been categorized or stratified.}
\begin{tabular}{@{}lccc@{}}
\toprule
                                     & Training  & Validation & Ratio ~ \\
\midrule
~ Medical material                   &     5818  &       313  & 0.053 ~ \\
~ Arterial wall calcification        &    13377  &       867  & 0.064 ~ \\
~ Cardiomegaly                       &     5308  &       325  & 0.061 ~ \\
~ Pericardial effusion               &     3412  &       226  & 0.066 ~ \\
~ Coronary artery wall calcification &    12025  &       765  & 0.063 ~ \\
~ Hiatal hernia                      &     6751  &       417  & 0.061 ~ \\
~ Lymphadenopathy                    &    12221  &       789  & 0.064 ~ \\
~ Emphysema                          &     9122  &       600  & 0.065 ~ \\
~ Atelectasis                        &    12263  &       713  & 0.058 ~ \\
~ Lung nodule                        &    21382  &      1361  & 0.063 ~ \\
~ Lung opacity                       &    17420  &      1184  & 0.067 ~ \\
~ Pulmonary fibrotic sequela         &    12589  &       831  & 0.066 ~ \\
~ Pleural effusion                   &     5705  &       376  & 0.065 ~ \\
~ Mosaic attenuation pattern         &     3638  &       253  & 0.069 ~ \\
~ Peribronchial thickening           &     4973  &       355  & 0.071 ~ \\
~ Consolidation                      &     8319  &       581  & 0.069 ~ \\
~ Bronchiectasis                     &     4732  &       330  & 0.069 ~ \\
~ Interlobular septal thickening     &     3745  &       249  & 0.066 ~ \\
\bottomrule
\end{tabular}
\label{tab:label-dist}
\end{table}

While CT scans provide comprehensive visibility of thoracic abnormalities, several of these conditions can also be discerned on chest X-rays, albeit with reduced clarity. Conditions such as cardiomegaly and pericardial effusion are typically indicated by alterations in the cardiac silhouette observable on X-rays. Hiatal hernia, if pronounced, lymphadenopathy in specific regions, and emphysema, inferred from hyperinflated lungs, may also be detected. Atelectasis, presenting as localized darkening due to reduced lung volume, lung nodules depending on their size and location, various lung opacities, and pulmonary fibrotic sequelae manifesting as scarring or lung thickening are visible under X-ray examination. Fluid accumulations like pleural effusion appear as a meniscus or diffuse haziness. Pulmonary consolidation, characterized by alveolar fluid, is evident as homogeneously dense areas. In severe cases, bronchiectasis can be indicated by noticeable airway dilatations, and interlobular septal thickening may show as prominent Kerley B lines. Conversely, features such as medical material, coronary artery wall calcification, and patterns specific to CT imaging like mosaic attenuation and peribronchial thickening are less discernible on X-ray, underscoring the complementary roles of X-ray and CT imaging in thoracic diagnosis.

Several labels overlap with existing large-scale chest X-ray datasets, notably CheXpert and NIH ChestXray14, including Cardiomegaly, Atelectasis, Lung Nodule, Lung Opacity, Pleural Effusion, and Consolidation. These will be the primary focus for the DRR-RATE dataset.

\section{Experiments and Results}
\label{sec:exp}

DRR-RATE's performance on the Chest X-ray classification task has been validated using CheXnet, a popular CXR classification network, along with the CheXpert dataset. CheXnet is kept at its default parameters, ensuring consistency in our experiments and allowing for direct comparisons with other studies using the same settings. All experiments are conducted on a Ubuntu 22.04.4 workstation, equipped with an Intel Xeon Gold 6130 CPU and an Nvidia A6000 GPU. All network hyperparameters are set to their default values or those commonly used in the literature. Code and results for the experiments are available at \url{https://github.com/farrell236/DRR-RATE}.

\begin{figure}[!ht]
\centering
\includegraphics[height=3cm]{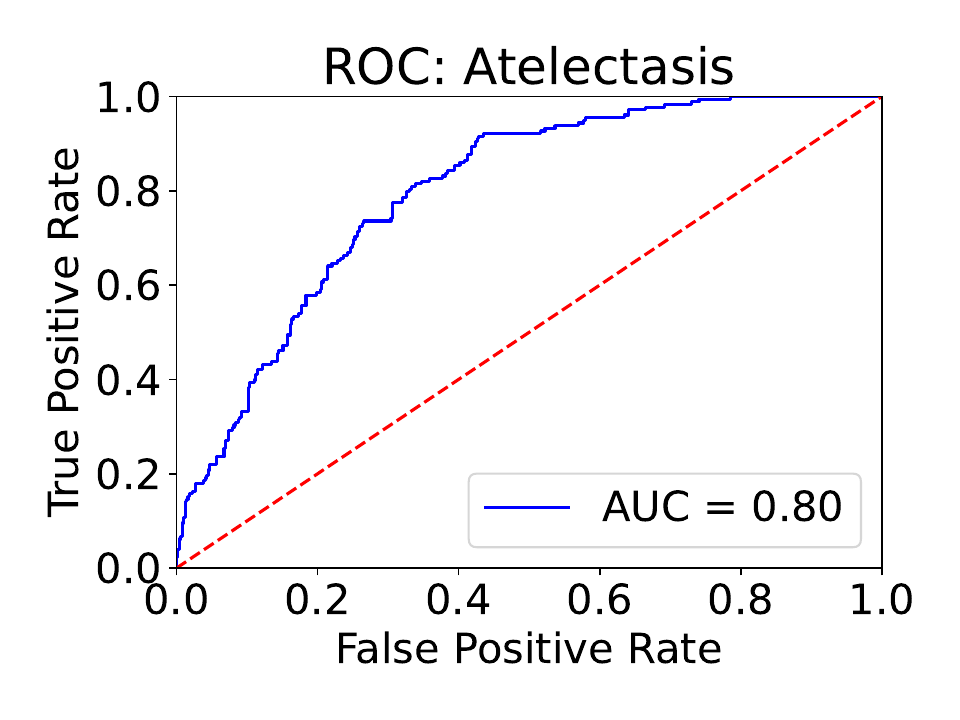}
\includegraphics[height=3cm]{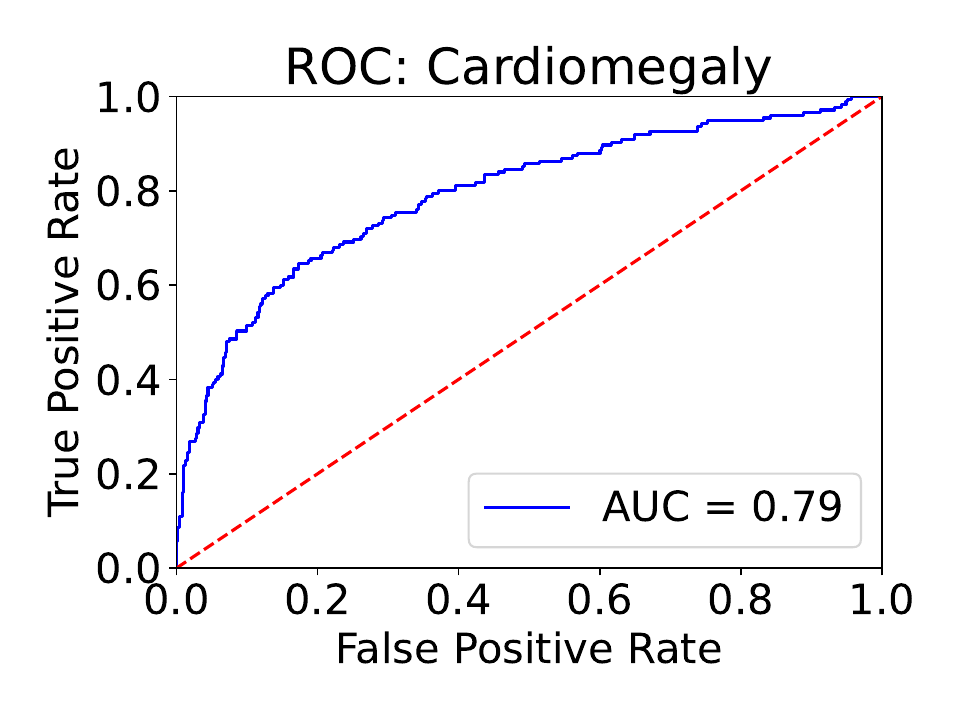}
\includegraphics[height=3cm]{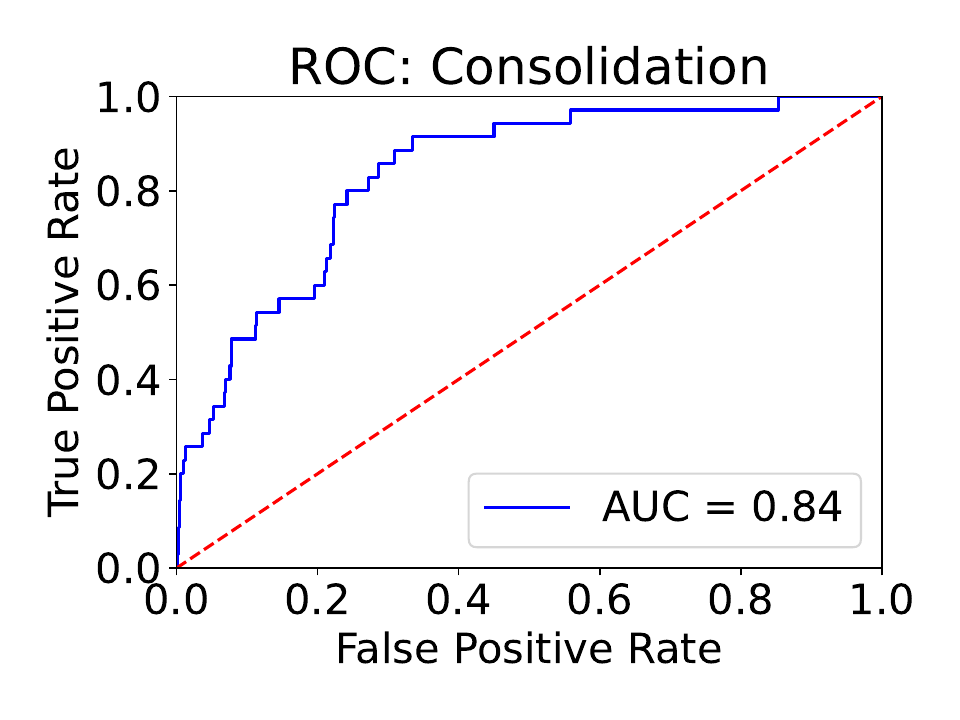} \\
\includegraphics[height=3cm]{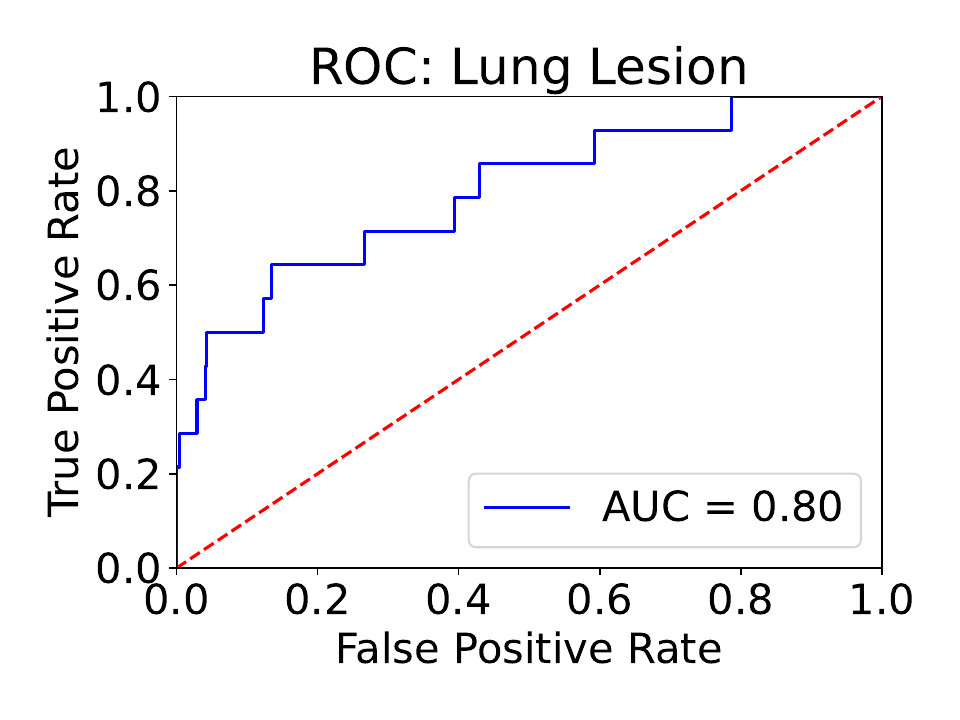}
\includegraphics[height=3cm]{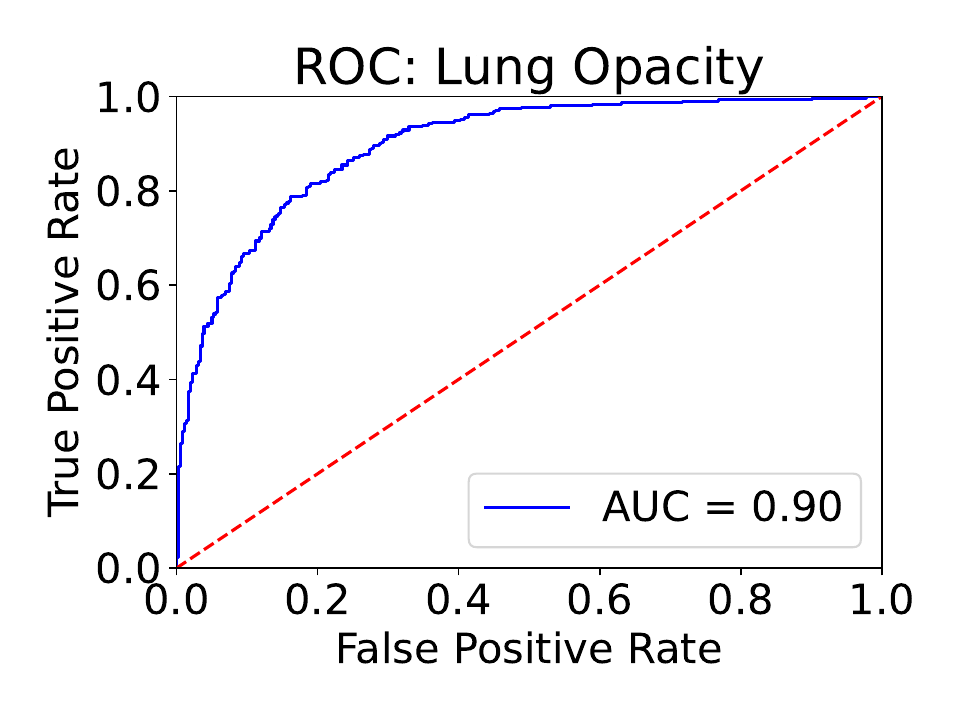}
\includegraphics[height=3cm]{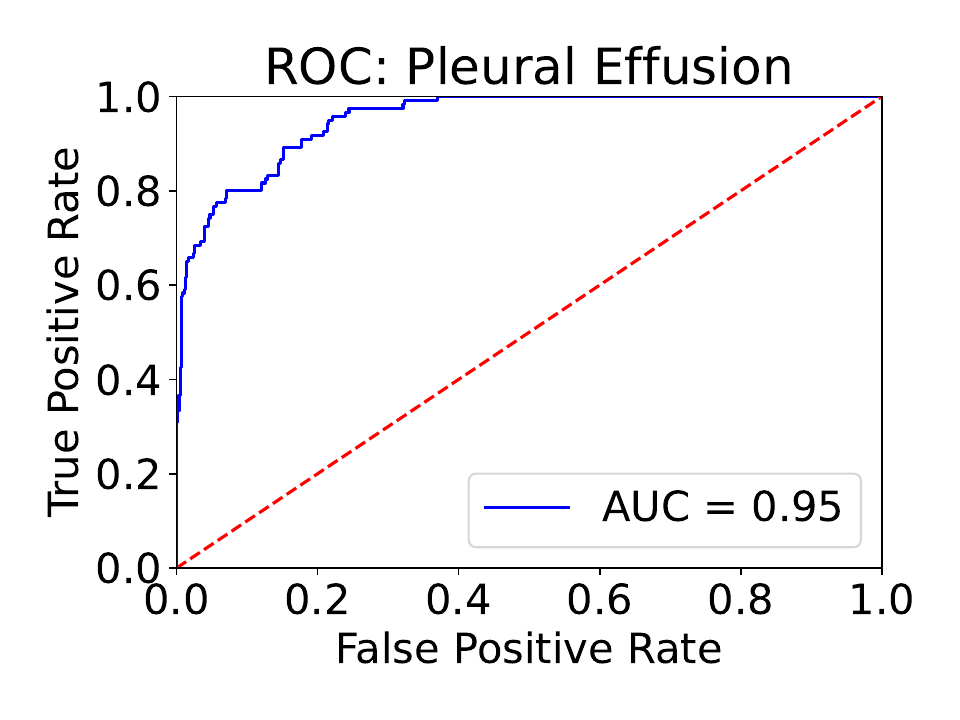}
\caption{ROC curve for CheXnet trained on CheXpert dataset. ROC scores = \{Atelectasis: 0.80, Cardiomegaly: 0.79, Consolidation: 0.84, Lung Lesion: 0.80, Lung Opacity: 0.90, and Pleural Effusion: 0.95\}.}
\label{fig:chexnet-chexpert}
\end{figure}

A baseline is established using the CheXpert dataset, which is divided into default train, validation, and test subsets with 223,415, 235, and 668 image-label pairs, respectively. A single CheXnet network was trained on the entire training set for a maximum of 50 epochs using the Adam optimizer, with a learning rate of 1e-4 and a weight decay of 1e-5. The weights were saved based on the best validation performance. The network reached convergence at epoch 22. Figure \ref{fig:chexnet-chexpert} shows the Receiver Operating Characteristic (ROC) curve for the six overlapping classes.

\begin{figure}[!ht]
\centering
\includegraphics[height=3cm]{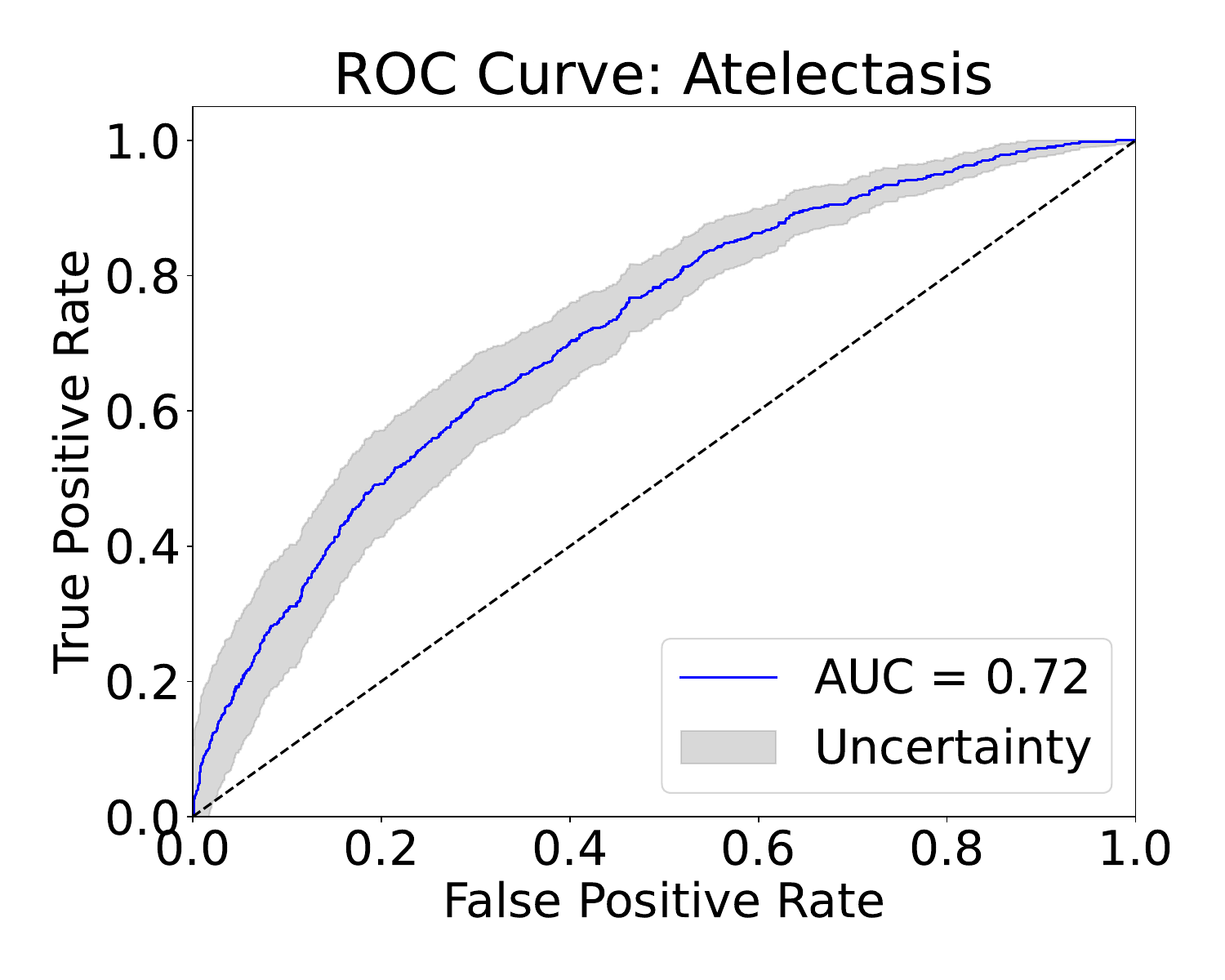}
\includegraphics[height=3cm]{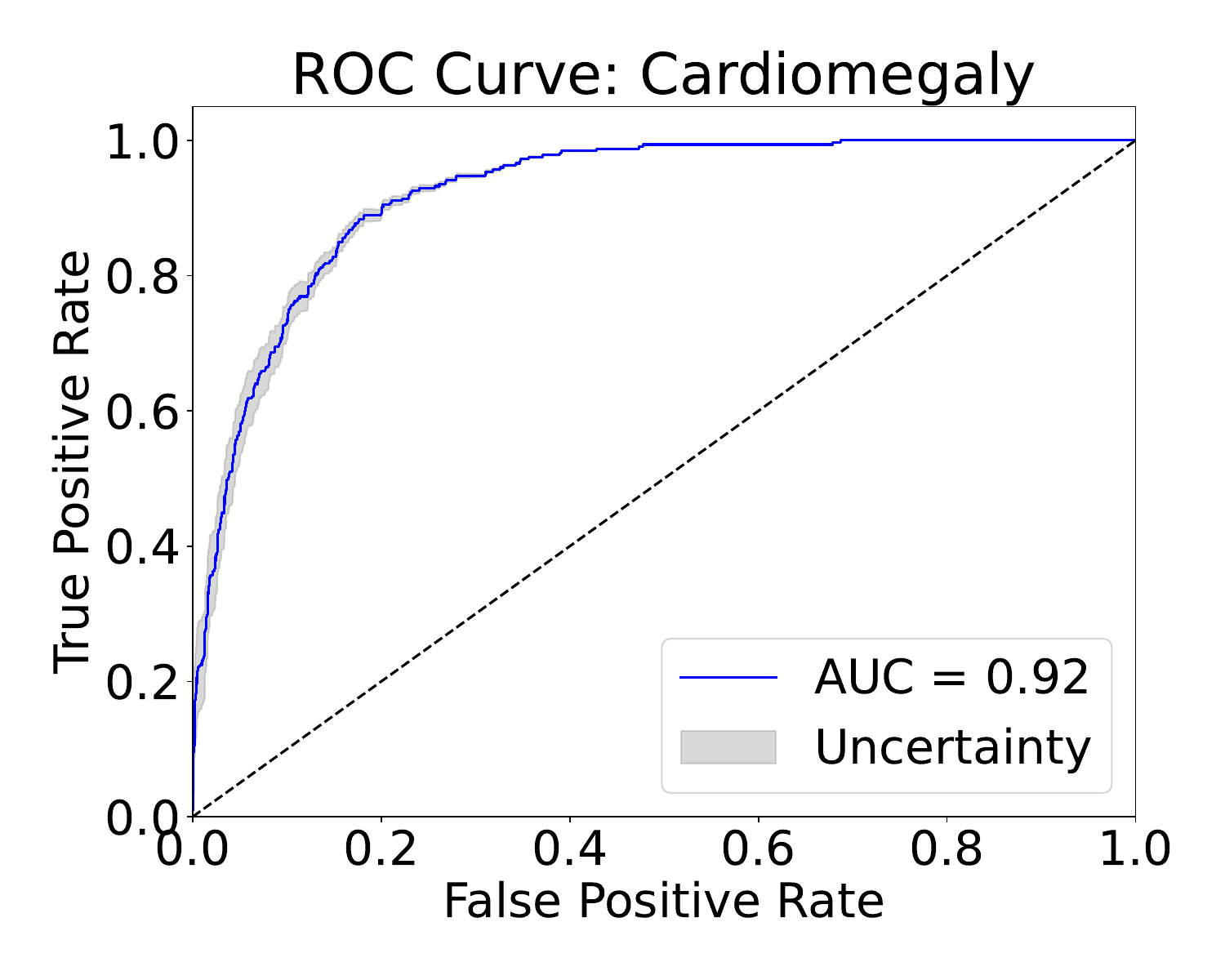}
\includegraphics[height=3cm]{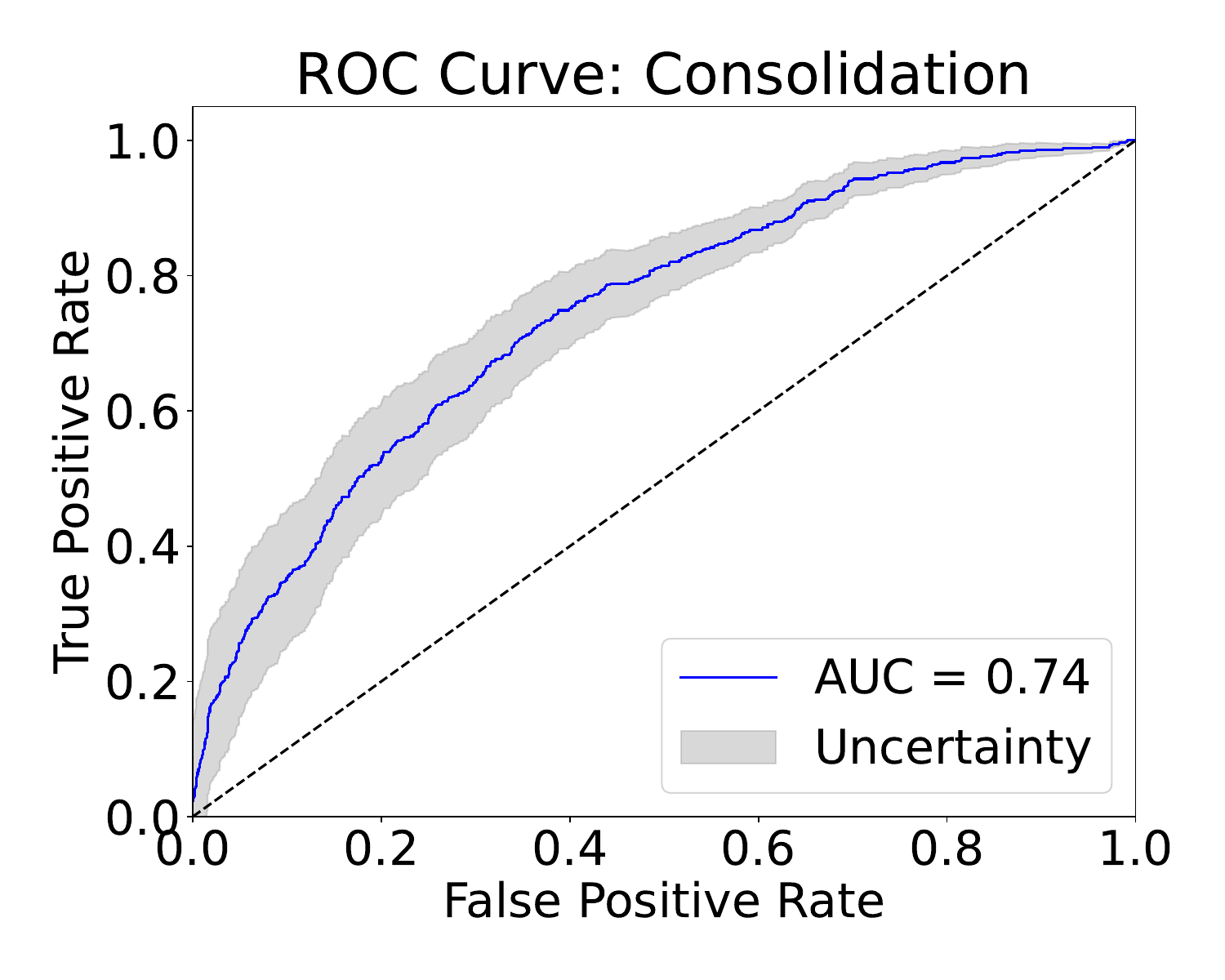} \\
\includegraphics[height=3cm]{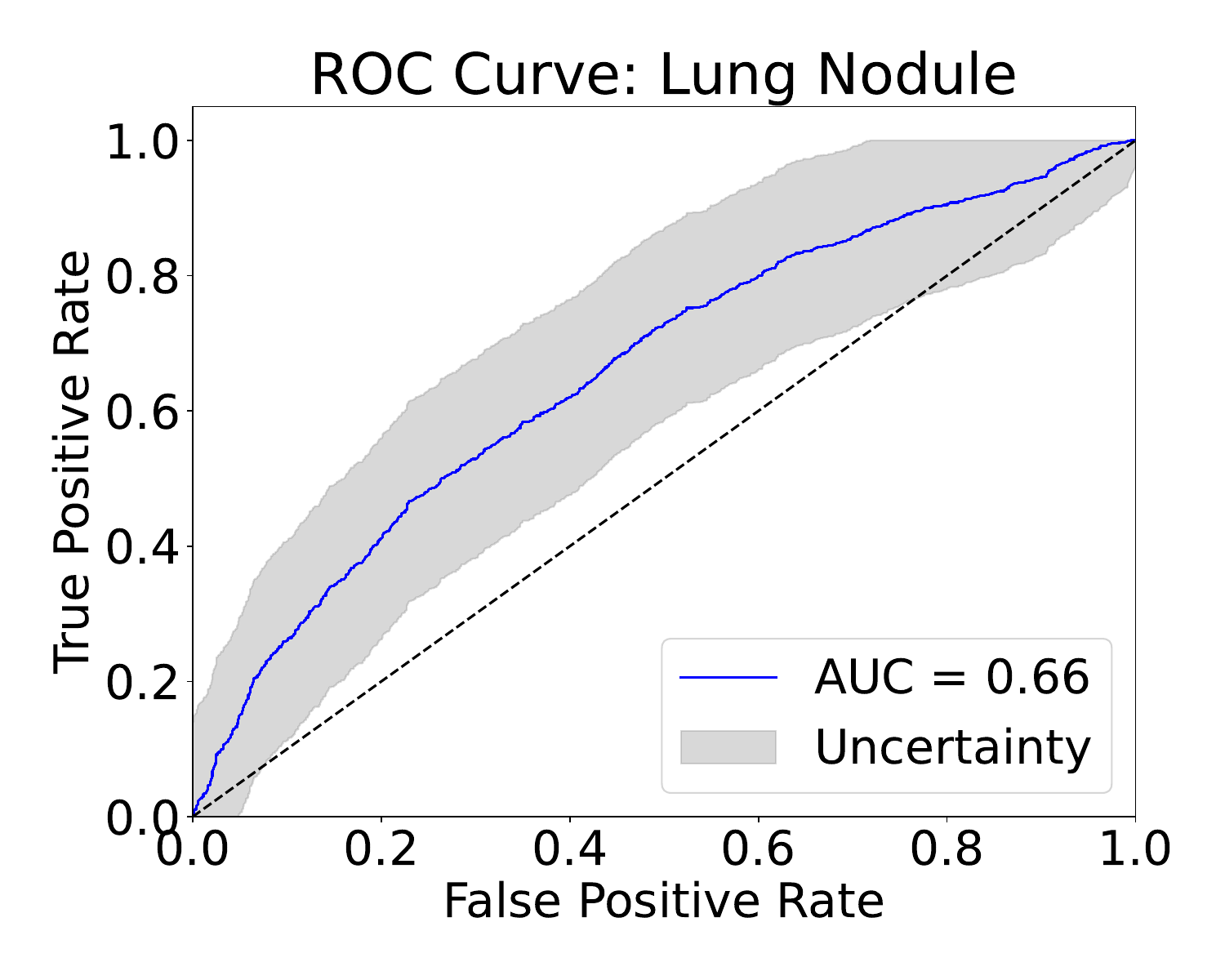}
\includegraphics[height=3cm]{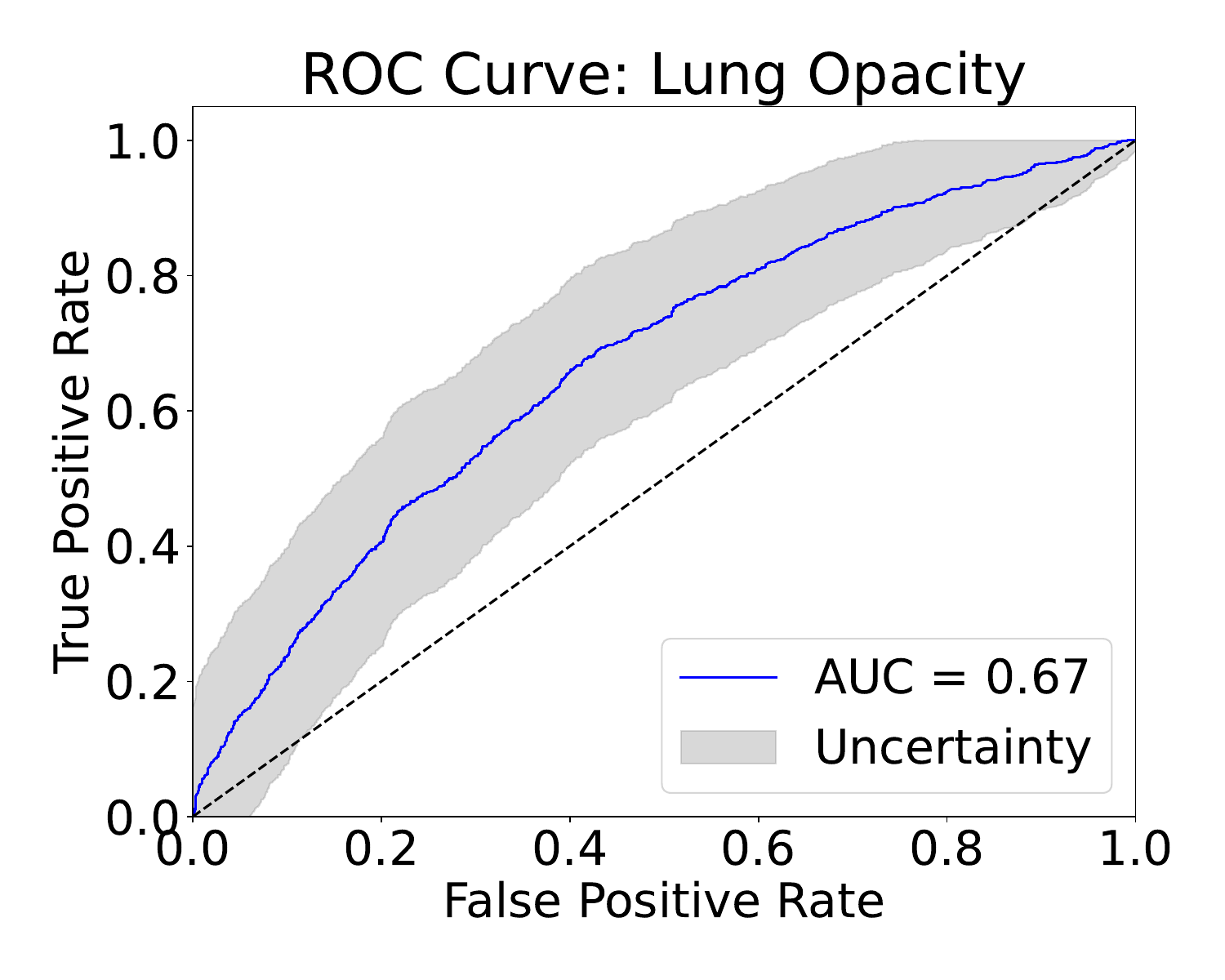}
\includegraphics[height=3cm]{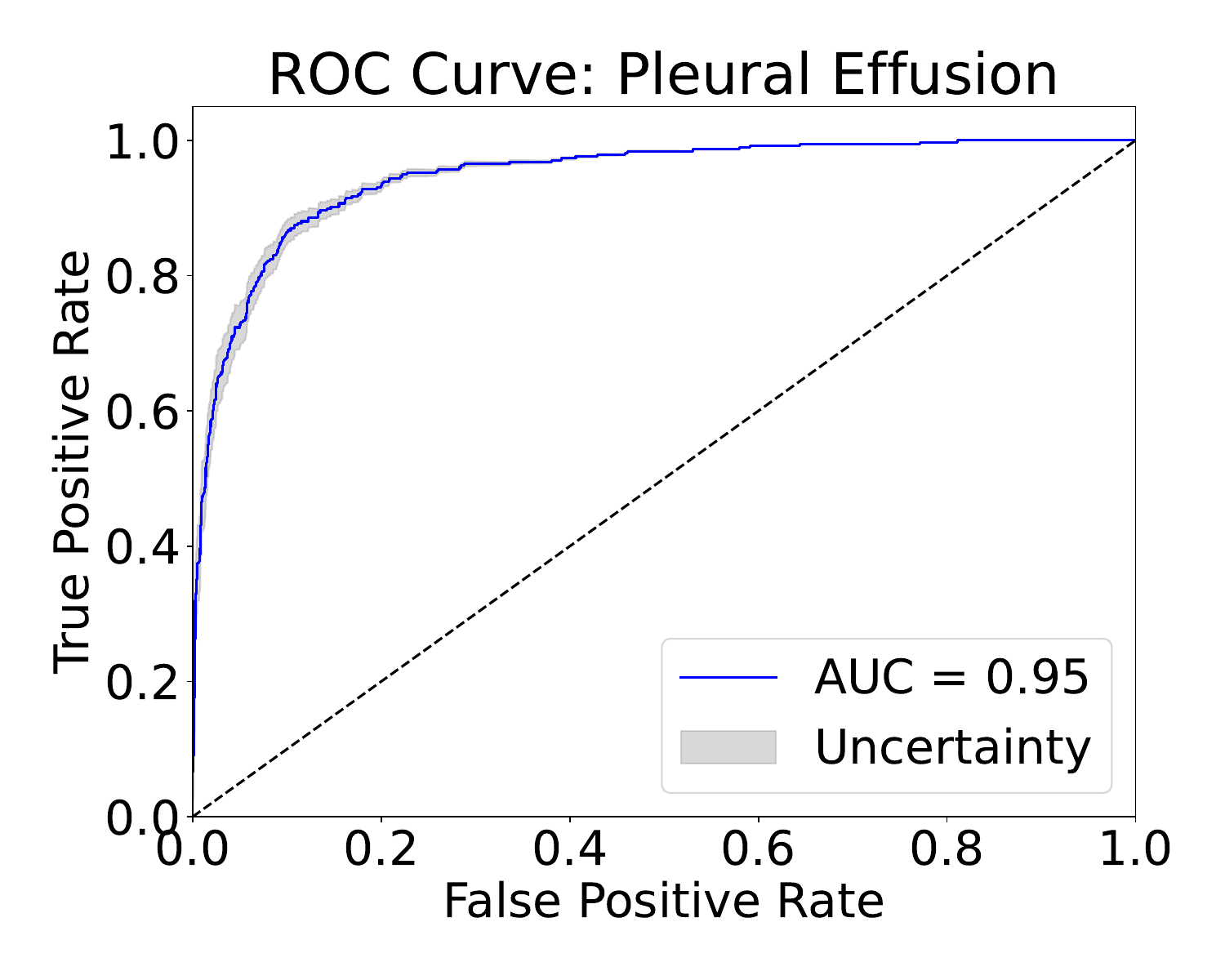}
\caption{ROC curve for CheXnet trained on DRR-RATE dataset. ROC scores = \{Atelectasis: 0.72, Cardiomegaly: 0.92, Consolidation: 0.74, Lung Nodule: 0.66, Lung Opacity: 0.67, and Pleural Effusion: 0.95\}. Grey shaded region denotes standard deviations at thresholds.}
\label{fig:chexnet-drr-rate}
\end{figure}

Similarly, CheXnet was trained on the DRR-RATE train and validation datasets, adhering to the default train and validation split used in CT-RATE. Since there is no official dedicated test set, five models were trained using five-fold cross-validation. CheXnet was trained for a maximum of 20 epochs using the Adam optimizer, with a learning rate of 1e-4 and a weight decay of 1e-5. The weights were saved based on the best validation performance, with the network reaching convergence between epochs 15 and 17. Figure \ref{fig:chexnet-drr-rate} illustrates the ROC curve for the six overlapping classes. 

Cardiomegaly and Pleural Effusion exhibit high AUC scores of 0.92 and 0.95 respectively, indicating superior predictive performance and reliability in identifying these conditions. On the other hand, Atelectasis and Consolidation show moderate AUC values of 0.72 and 0.74, suggesting decent, albeit less consistent, performance. The classes of Lung Nodule and Lung Opacity have the lowest AUC scores, at 0.66 and 0.67 respectively, pointing to a need for improvement in model accuracy for these conditions. The uncertainty, depicted by the grey shaded regions, is noticeably smaller for Cardiomegaly and Pleural Effusion, indicating less variation in true positive rates at different thresholds and hence more robust model predictions. In contrast, the broader uncertainty bands for conditions like Lung Nodule and Lung Opacity reflect greater variability in model performance.

Comparing the ROC curves of the CheXnet model trained on the DRR-RATE dataset, with those from the CheXnet model trained on the CheXpert dataset, reveals generally comparable performance for Atelectasis, Cardiomegaly, Consolidation, and Pleural Effusion. However, the performance for Lung Nodule and Lung Opacity is slightly reduced.

\begin{figure}[!ht]
\centering
\includegraphics[height=3cm]{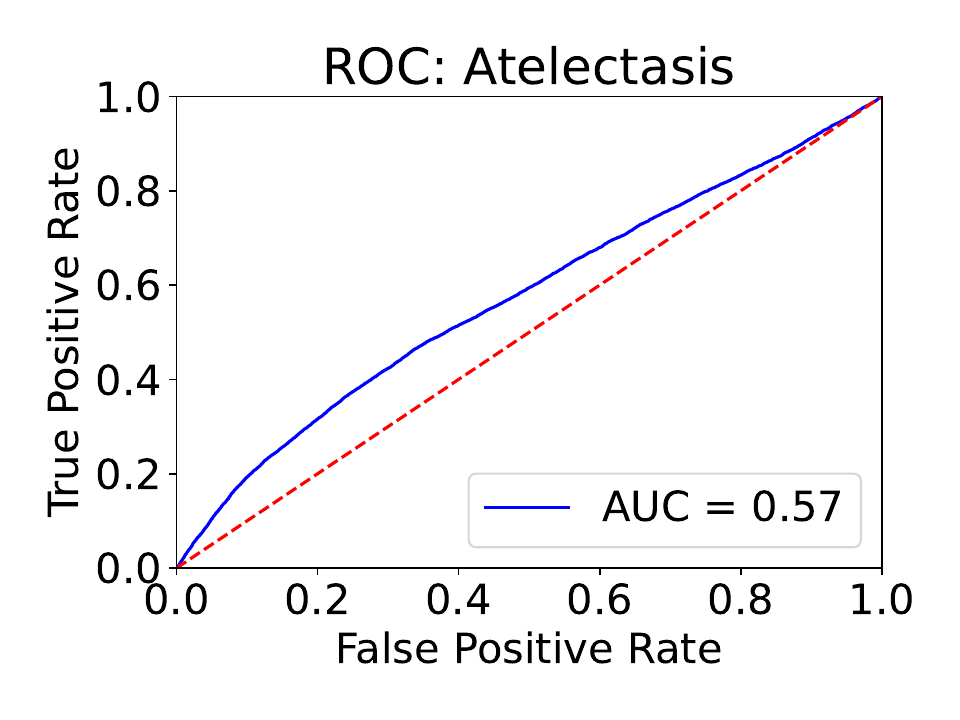}
\includegraphics[height=3cm]{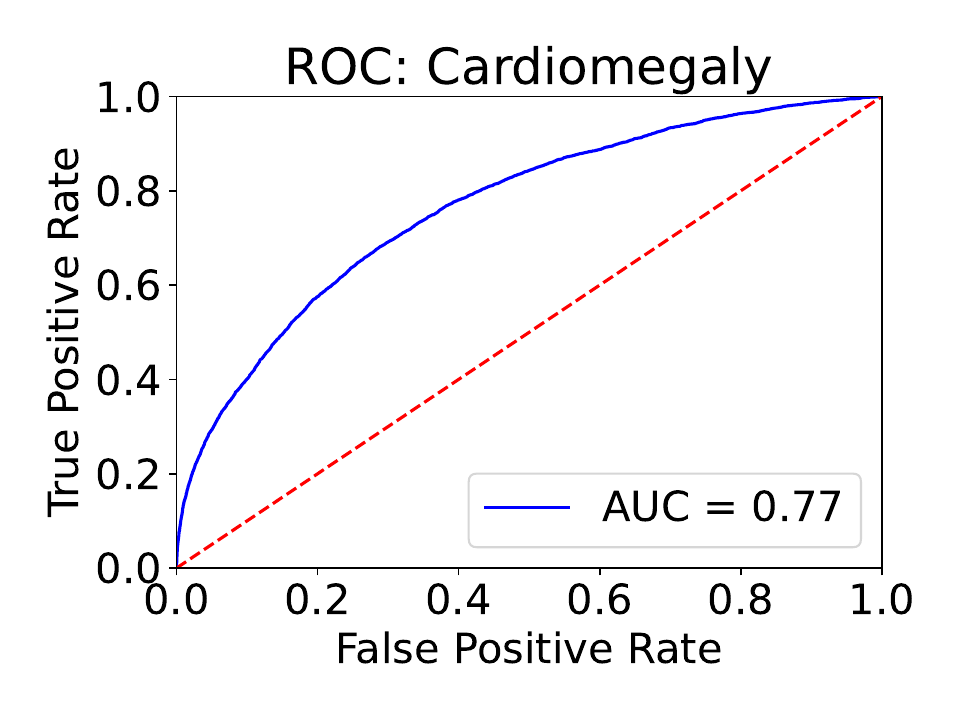}
\includegraphics[height=3cm]{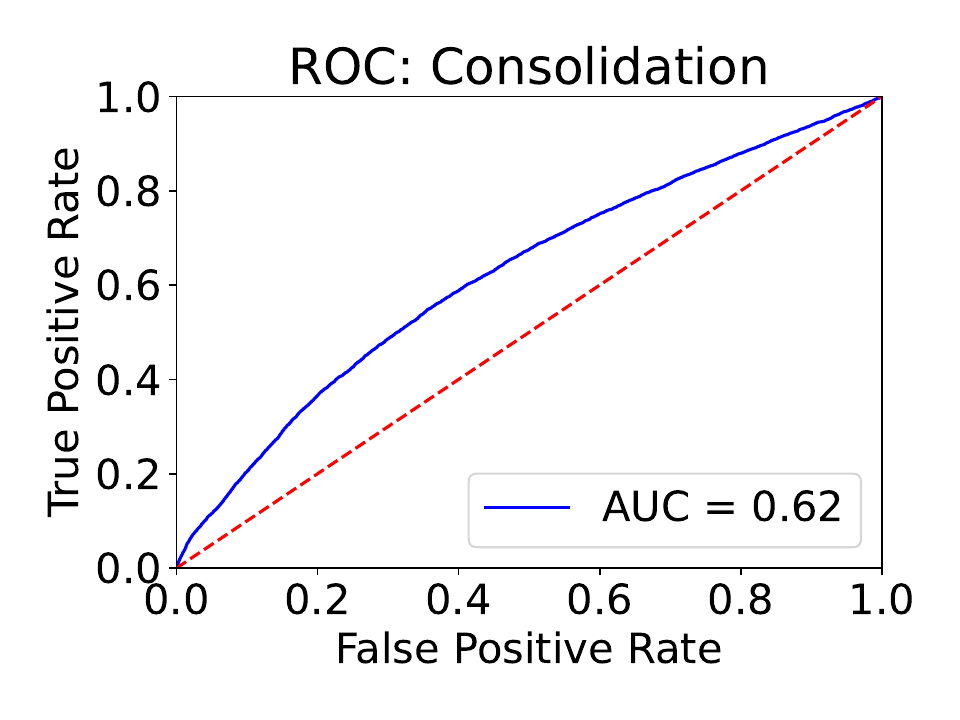} \\
\includegraphics[height=3cm]{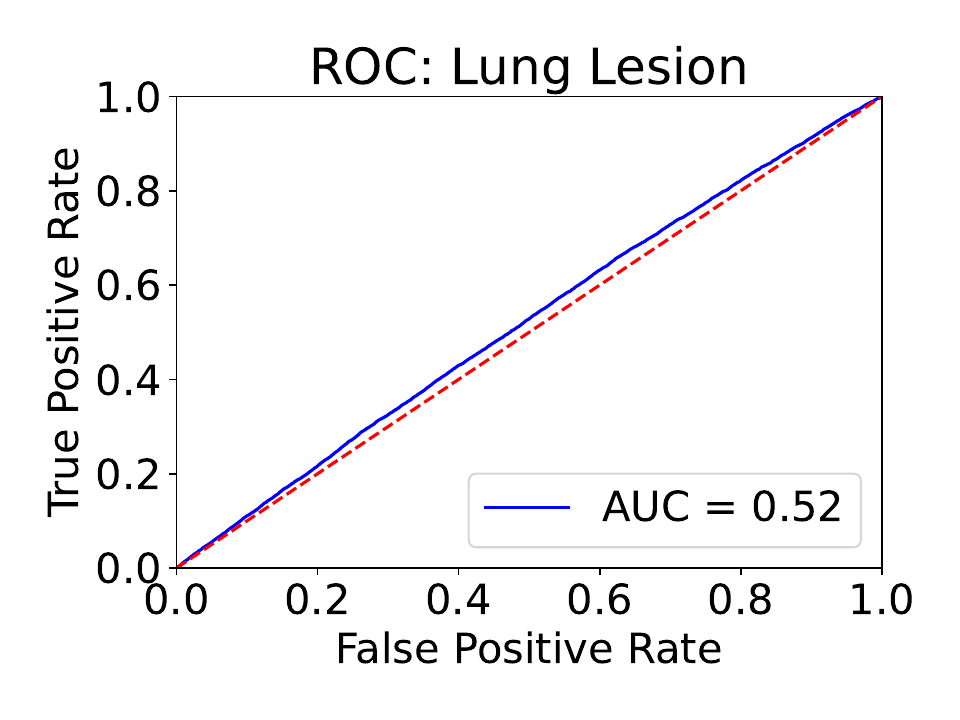}
\includegraphics[height=3cm]{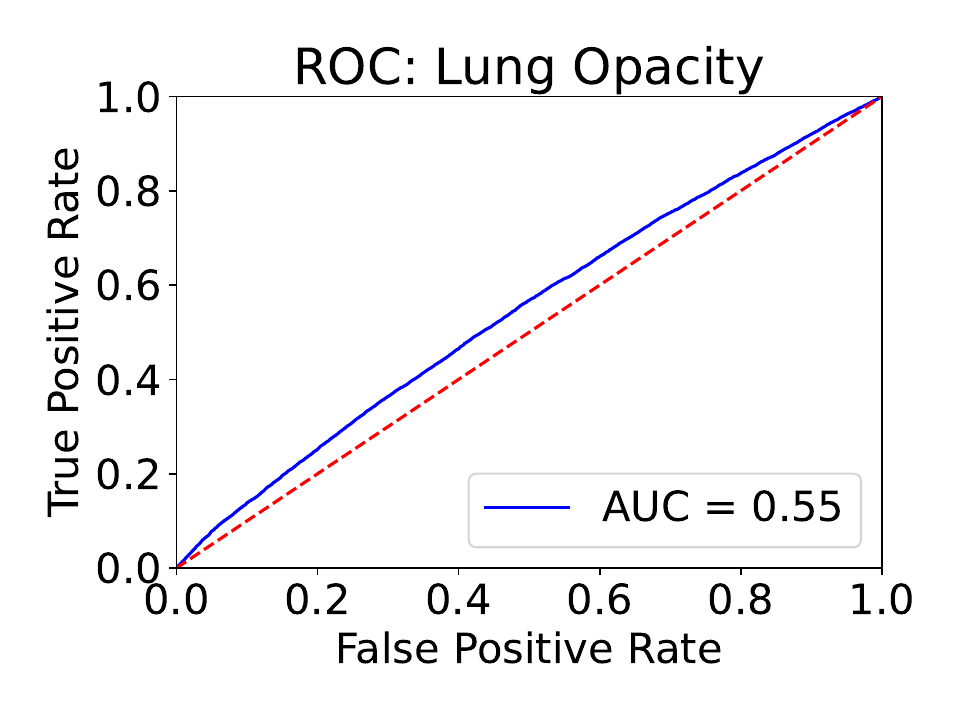}
\includegraphics[height=3cm]{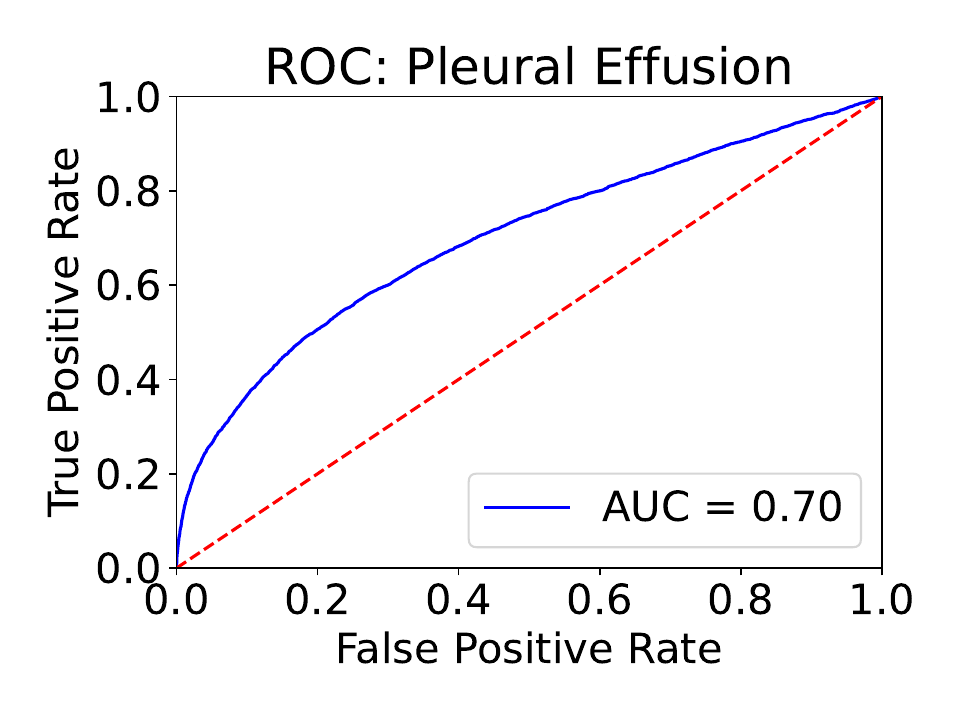}
\caption{ROC curve for CheXnet trained on CheXpert dataset and test on DRR-RATE dataset. ROC scores = \{Atelectasis: 0.57, Cardiomegaly: 0.77, Consolidation: 0.62, Lung Nodule: 0.52, Lung Opacity: 0.55, and Pleural Effusion: 0.70\}}
\label{fig:chexpert-drr-rate}
\end{figure}

Figure \ref{fig:chexpert-drr-rate} presents ROC curves for a CheXnet model trained on the CheXpert dataset and tested on the DRR-RATE dataset. Cardiomegaly and Pleural Effusion maintained AUC scores of 0.77 and 0.70, respectively, while Consolidation achieved a score of 0.62. Atelectasis, Lung Lesion, and Lung Opacity scored between 0.50 and 0.60, indicating performance slightly above random guessing. This decrease is likely due to domain shift between ``real'' and DRR images. 

\section{Discussion and Conclusion}
\label{sec:discussion}
In this paper, we introduced DRR-RATE, a large synthetic chest X-ray dataset complete with labels and radiological reports. We have demonstrated that synthesizing images from CT scans is an effective method for capturing pathologies typically present in CT images. This approach not only enriches the data available for training diagnostic models but also enhances our understanding of how these pathologies manifest in different imaging modalities.

A baseline CheXnet model was trained on both the CheXpert dataset and the DRR-RATE dataset, and tested each model on its respective dataset. Both models demonstrated high AUC scores. Additionally, when testing the CheXnet model that was trained on the CheXpert dataset to evaluate its effectiveness on synthetic DRR images, it performed best in detecting Cardiomegaly, Consolidation, and Pleural Effusion. These pathologies are typically more prominent and have more distinct features, which likely contributed to the higher accuracy. However, the model was less effective in identifying Atelectasis, Lung Nodule, and Lung Opacity. Certain findings, such as Lung Nodule, Lung Opacity, and Atelectasis, may be small on CT scans, and if they are minor, they might not be visible on the DRR images. Additionally, the definition of these conditions may vary, as different institutions often use distinct grading criteria.

The rendered DRR images have a resolution of 512 $\times$ 512 pixels, which may affect the visibility of smaller features, especially when compared to digital radiographs that typically exceed 2000 $\times$ 2000 pixels in size. While it is feasible to create higher-resolution DRRs, the resolution of the underlying CT scans also plays a critical role. If the CT resolution is low, generating very high-resolution DRRs (i.e., oversampling) may not be necessary. Moreover, it is common in the literature for chest X-ray images to be downsampled to 224 $\times$ 224 pixels before being processed by neural networks, which may reduce the necessity for creating high-resolution DRR images. In the production of DRRs, scattering and refraction of X-rays are not considered, although the results are sufficiently close to real X-ray images for practical applications. Additionally, if the CT contains artifacts, such as metal artifacts, these scans are not suitable for DRR creation since the resulting pixels mirror the standard CXR image formation process.

The application of DRRs, particularly paired X-ray and CT data, is increasingly valuable in the current era dominated by vision-based large language models. Identifying findings in both chest and abdominal DRRs, as potentially producible from CT datasets like DeepLesion, offers significant benefits. For example, detecting abnormalities in localizer images before advancing to more detailed CT imaging can potentially eliminate the need for contrast agents if the preliminary findings are sufficiently clear. This capability not only streamlines diagnostic processes but also reduces patient exposure to additional chemicals and radiation.

The use of AI in chest X-ray diagnosis presents several risks, including biases, job displacement, and privacy concerns. Additionally, AI errors and unequal access could exacerbate healthcare disparities. It is crucial to utilize diverse training datasets, enhance privacy protections, educate healthcare professionals on AI integration. The introduction of DRR-RATE provides a new kind of paired data modality, enhancing current capabilities. 

In conclusion, the development and application of DRR-RATE represent significant advances in the synthesis and use of medical imaging data. By effectively replicating key pathological features from CT scans in X-ray-like images, this dataset facilitates a wide range of research opportunities in machine learning and medical diagnostics. 

\section*{Acknowledgements}

This work was supported by the Intramural Research Program of the NIH Clinical Center and National Library of Medicine. This work utilized the computational resources of the NIH HPC Biowulf cluster. (\url{http://hpc.nih.gov})

\bibliographystyle{ieeetr}
\bibliography{references}

\newpage
\section*{Checklist}


\begin{enumerate}

\item For all authors...
\begin{enumerate}
  \item Do the main claims made in the abstract and introduction accurately reflect the paper's contributions and scope?
    \answerYes{}
  \item Did you describe the limitations of your work?
    \answerYes{See Section \ref{sec:discussion}.}
  \item Did you discuss any potential negative societal impacts of your work?
    \answerYes{See Section \ref{sec:discussion}.}
  \item Have you read the ethics review guidelines and ensured that your paper conforms to them?
    \answerYes{}
\end{enumerate}

\item If you are including theoretical results...
\begin{enumerate}
  \item Did you state the full set of assumptions of all theoretical results?
    \answerNA{}
	\item Did you include complete proofs of all theoretical results?
    \answerNA{}
\end{enumerate}

\item If you ran experiments (e.g. for benchmarks)...
\begin{enumerate}
  \item Did you include the code, data, and instructions needed to reproduce the main experimental results (either in the supplemental material or as a URL)?
    \answerYes{Experiment Code accessible via \url{https://github.com/farrell236/DRR-RATE}}
  \item Did you specify all the training details (e.g., data splits, hyperparameters, how they were chosen)?
    \answerYes{DRR Generation Parameters (Section \ref{sec:siddonjacobs}): Most parameters were kept at their default settings. Data Split (Section \ref{sec:exp}): The default train-to-validation split used in CT-RATE was adhered to. Network Hyperparameters (Section \ref{sec:exp}): All network hyperparameters were set to their default values or those commonly used in the literature.}
	\item Did you report error bars (e.g., with respect to the random seed after running experiments multiple times)?
    \answerYes{In Section \ref{sec:exp}, we conduct cross-validation training, report the model's uncertainty, and include the set seed in the experiment code.}
	\item Did you include the total amount of compute and the type of resources used (e.g., type of GPUs, internal cluster, or cloud provider)?
    \answerYes{All experiments are conducted on a Ubuntu 22.04.4 workstation, equipped with an Intel Xeon Gold 6130 CPU and an Nvidia A6000 GPU}
\end{enumerate}

\item If you are using existing assets (e.g., code, data, models) or curating/releasing new assets...
\begin{enumerate}
  \item If your work uses existing assets, did you cite the creators?
    \answerYes{CT-RATE Dataset detailed in ``A foundation model utilizing chest CT volumes and radiology reports for supervised-level zero-shot detection of abnormalities'' by Hamamci et al. \cite{hamamci2024foundation}}
  \item Did you mention the license of the assets?
    \answerYes{See Section \ref{sec:drr-rate},}
  \item Did you include any new assets either in the supplemental material or as a URL?
    \answerYes{The dataset is publicly accessible at \url{https://huggingface.co/datasets/farrell236/DRR-RATE}}
  \item Did you discuss whether and how consent was obtained from people whose data you're using/curating?
    \answerNo{The data has been anonymized, and therefore, informed consent is not required.}
  \item Did you discuss whether the data you are using/curating contains personally identifiable information or offensive content?
    \answerNo{The data was previously anonymized and contains no PII.}
\end{enumerate}

\item If you used crowdsourcing or conducted research with human subjects...
\begin{enumerate}
  \item Did you include the full text of instructions given to participants and screenshots, if applicable?
    \answerNA{}
  \item Did you describe any potential participant risks, with links to Institutional Review Board (IRB) approvals, if applicable?
    \answerNA{}
  \item Did you include the estimated hourly wage paid to participants and the total amount spent on participant compensation?
    \answerNA{}
\end{enumerate}

\end{enumerate}

\end{document}